\documentclass[aps,prl,twocolumn,superscriptaddress,showpacs]{revtex4}
\usepackage{amsmath}
\usepackage{dcolumn}
\usepackage{epsfig}
\usepackage{graphicx}
\usepackage{latexsym}
\usepackage{color}

\def\LBCO{La$_{2-x}$Ba$_x$CuO$_4$}
\def\LBBCO{La$_{1.875}$Ba$_{0.125}$CuO$_4$}
\def\LSCO{La$_{2-x}$Sr$_x$CuO$_4$}
\def\YBCO{YBa$_2$Cu$_3$O$_{6+x}$}

\def\BSCO{Pb$_{0.55}$Bi$_{1.5}$Sr$_{1.6}$La$_{0.4}$CuO$_{6+\delta}$}

\begin{document}

%\twocolumn[
%\hsize\textwidth\columnwidth\hsize\csname@twocolumnfalse\endcsname

%\draft

\title{Magneto-optical signatures of a cascade of transitions in La$_{1.875}$Ba$_{0.125}$CuO$_4$}

\author{Hovnatan Karapetyan}
\affiliation{Stanford Institute for Materials and Energy Sciences, SLAC National Accelerator Laboratory, 2575 Sand Hill Road, Menlo Park, CA 94025, USA}
\affiliation{Department of Applied Physics, Stanford University, Stanford, CA 94305, USA}
\author{M. H\"ucker}
\affiliation{Condensed Matter Physics and Materials Science Department, Brookhaven National Laboratory, Upton, NY 11973-5000, USA}
\author{G. D. Gu}
\affiliation{Condensed Matter Physics and Materials Science Department, Brookhaven National Laboratory, Upton, NY 11973-5000, USA}
\author{J. M. Tranquada}
\affiliation{Condensed Matter Physics and Materials Science Department, Brookhaven National Laboratory, Upton, NY 11973-5000, USA}
\author{M. M. Fejer}
\affiliation{Department of Applied Physics, Stanford University, Stanford, CA 94305, USA}
\author{Jing Xia}
\affiliation{Department of Physics and Astronomy, University of California, Irvine, CA 92697-4575, USA}
\author{A. Kapitulnik}
\affiliation{Stanford Institute for Materials and Energy Sciences, SLAC National Accelerator Laboratory, 2575 Sand Hill Road, Menlo Park, CA 94025, USA}
\affiliation{Department of Applied Physics, Stanford University, Stanford, CA 94305, USA}
\affiliation{Department of Physics, Stanford University, Stanford, CA 94305, USA}
%\email{aharonk@stanford.edu}

\date{\today}

\begin{abstract}
Recent  experiments on the original cuprate high temperature superconductor, \LBCO, revealed a remarkable sequence of phase transitions \cite{qli1}. Here we investigate such crystals with polar Kerr effect which is sensitive to time-reversal-symmetry breaking. Concurrent birefringence measurements accurately locate the structural phase transitions from high-temperature tetragonal to low temperature orthorhombic, and then to lower temperature tetragonal, at which temperature a strong Kerr signal onsets. Hysteretic behavior of the Kerr signal suggests that time-reversal symmetry is already broken well above room temperature, an effect that was previously observed in high quality \YBCO~crystals \cite{xia1}. 
\end{abstract}

\pacs{74.25.Bt, 74.25.Gz, 74.72.Kf, 75.30.Fv}

\maketitle

Of the known high-temperature superconductors (HTSC), \LBCO, and in particular where $x=0.125$, has provided invaluable information on the interplay between superconductivity and other competing phases. A deep depression in the superconducting phase boundary [$T_{\mathrm{c}}(x)$], centered at $x = 1/8$ \cite{moodenbaugh,yamada,qli1}, reveals  structural phase transitions and charge and spin stripe ordering \cite{hucker2} that appear static below $\sim 50$~K. Despite the anticipated competition with global superconductivity, recent work on this system has provided evidence for the development of strong two-dimensional (2D) superconducting correlations for $T\lesssim 40$~K \cite{qli1,tranquada1}, suggesting  that stripe order may not directly compete with pairing correlations within the CuO$_2$ planes, but instead frustrates the Josephson coupling between layers \cite{berg3,tajima}, thus inhibiting the development of 3D superconducting order. 

Early studies of \LBCO~near $x=1/8$ revealed a sequence of structural phase transitions, first  from a high-temperature tetragonal (HTT) phase to a low-temperature orthorhombic (LTO) phase at $T_{\mathrm{HT}}\approx 230$~K, followed by a LTO to LTT (low-temperature tetragonal) at  $T_{\mathrm{LT}}\approx 54$~K \cite{axe,billinge}. Magnetic correlations found  below $T_{\mathrm{LT}}$ were first interpreted as antiferromagnetic (AF) order \cite{kumagai,luke,sera,arai}, while in more recent studies, localized Cu moments were found to dominate the magnetic response \cite{hucker1}, suggesting that  charge and spin stripes  are being formed  as a way for local-AF spin correlations to coexist with mobile holes in the doped cuprates \cite{tranquada1,zaanen,kivelson1,voita}. 

The existence of magnetic correlations raises the question of whether time reversal symmetry (TRS) is broken at any temperature in the phase diagram of \LBBCO,  whether it is related to the structural and stripe-ordered phases observed so far, and whether it has any relation to superconductivity which appears at lower temperatures. Indeed, TRS-breaking (TRSB) effects have been predicted \cite{varma,chakravarty,berg3}, and observed \cite{fauque,xia1,rhhe,yli} in the pseudogap phase of HTSC, with the recent observation of a possible strong connection between the occurrence of TRSB and some kind of charge ordering (CO) in single-layer \BSCO~cuprate (BSCO) \cite{rhhe}.

\begin{figure}[h]
\begin{center}
\includegraphics[width=1.0 \columnwidth]{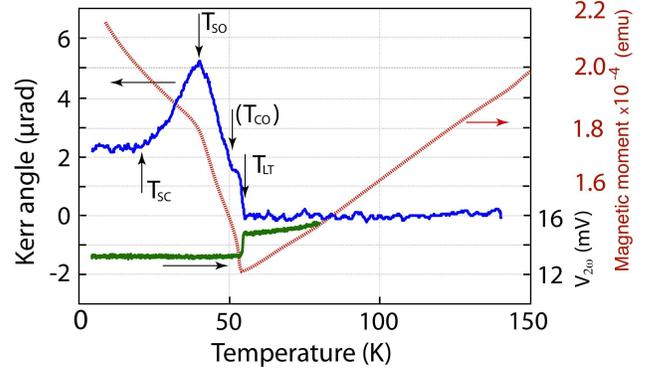}
\end{center}
\vspace{-5mm}
\caption{ Zero-field cool Kerr effect of a cleaved sample. Here we mark i) the LTO to LTT transition, $T_{\mathrm{LT}}$, which is also evident in the second-harmonic data; ii) the onset of CO, $T_{\mathrm{CO}}$; the onset of spin ordering, $T_{\mathrm{SO}}$; and the temperature below which superconductivity is established in the $a$-$b$ plane, $T_{\mathrm{SC}}$.  The location of $T_{\mathrm{CO}}$, $T_{\mathrm{SO}}$, and $T_{\mathrm{SC}}$ are taken from Li {\it et al.} \cite{qli1}. Susceptibility (dotted line) was measured on the same crystal at a magnetic field of 1~T applied in the $c$-direction, and is identical to the data in ref. \cite{hucker1}.} 
\label{features}
\end{figure}

In this letter we present optical-birefringence and magneto-optical (MO) data on \LBBCO~single crystals. The intensity of two circularly polarized light beams interfering at the detector, proportional to the change in the birefringence of the sample, accurately locate the structural phase transitions at $T_{\mathrm{HT}}$ and $T_{\mathrm{LT}}$, while a strong Kerr signal onsetting at $T_{\mathrm{LT}}$ indicates that TRS is evidently  {\it at least} broken below that temperature. As shown in Fig.~\ref{features}, the Kerr signal increases below $T_{\mathrm{LT}}$, followed by a weak inflection that possibly indicates the CO transition $T_{\mathrm{CO}}$ \cite {tranquada1} (it is usually found that $T_{\mathrm{LT}}\approx T_{\mathrm{CO}}$ for $x=1/8$  \cite{hucker1},) rises to a maximum around the spin-order transition $T_{\mathrm{SO}}$, and decreases to a finite value when superconducting correlations are substantial ($\sim 25$~K).  However, despite the sharp onset of the Kerr signal, hysteretic training effects are observed, indicating that TRS has been broken at much higher temperatures.  Similar effect was previously observed in \YBCO~(YBCO), especially close to $x=1/8$ \cite{xia1}, on single-layer BSCO \cite{rhhe}, and has recently reported on similar \LBBCO~crystals by Li {\it et al.} \cite{luli}. Our results, together with the detailed magnetic studies on similar crystals \cite{hucker1}, may point to a unique magnetic structure in the material that is strongly altered when CO occurs, so as to allow the Kerr effect to be visible. 

The crystals studied here were grown in an infrared image furnace by the floating-zone technique. Some are pieces from the same crystals used previously to characterize the optical conductivity \cite{homes1}, photoemission and scanning tunneling spectroscopy (STS) \cite{valla1}, magnetization \cite{hucker1}, and magnetic excitations \cite{tranquada2}. In particular, the charge-stripe order has been characterized previously by soft x-ray resonant diffraction \cite{abbamonte}. Six crystals have been studied to date, of which one large crystal was cleaved into several smaller pieces, and three were measured separately.  All crystals showed qualitatively similar behavior, but with different strength of the effect. While the structural phase transitions have been located to within 0.2~K in all crystals, the CO transition varied within 4~K below the LTO to LTT transition, and the SO transition varied over a wider range as discussed below.   In this paper we show data on the crystal with the strongest Kerr and birefringence response. Measurements were taken on numerous locations, spread all over the face of the cleaved crystal. Hard x-ray scattering studies of the electronic and lattice modulations associated with stripe order in this same crystal confirmed that it has the sharpest LTT transition of all the samples looked at by this technique. It also shows the strongest scattering both from the LTT peak and the CO peak. It is believed that this crystal was grown with minimal composition gradient and has no ``dead layers" at its high quality cleaved surface \cite{wilkins}. Other polished, or ``less clean" cleaved crystals showed a signal about five to eight times smaller, but with otherwise similar temperature dependencies.

\begin{figure}[h]
\begin{center}
\includegraphics[width=1.0 \columnwidth]{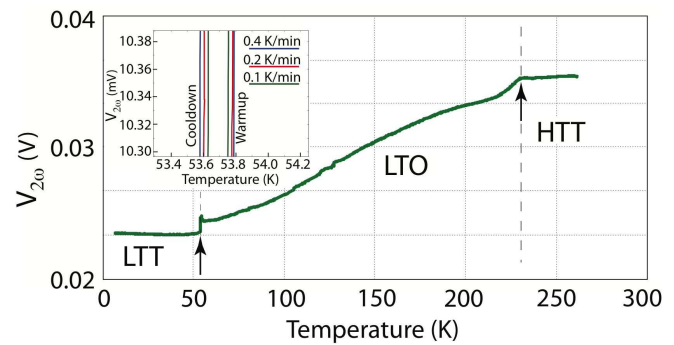}
\end{center}
\vspace{-5mm}
\caption{Second-harmonic signal of our apparatus detecting evolution of the birefringence in the $a$-$b$ plane of the sample. Arrows mark the second-order transition from HTT to LTO at $T_{\mathrm{HT}}=230.8$~K, and the first-order transition from LTO to LTT at $T_{\mathrm{LT}}=53.7$~K (see text). Inset shows a blow up of the middle of the first-order transition, taken at lower power, and at three different cycle rates. Clearly the higher the rate, the larger is the hysteresis in the birefringence of the sample.} 
\label{structural}
\end{figure}

MO effects are described within quantum theory as the interaction of photons with electron spins through spin-orbit interaction \cite{pershan}. Macroscopically, linearly polarized light that interacts with magnetized media can exhibit both ellipticity and a rotation of the polarization state. The leading terms in any MO effects are proportional to the off-diagonal part of the ac conductivity:  $\sigma_{xy}(\omega)= \sigma_{xy}^\prime(\omega) +i\sigma_{xy}^{\prime \prime}(\omega)$,  through the asymmetry between the complex indices of refraction for right and left circularly polarized light $(\tilde{n}_\mathrm{R} \neq \tilde{n}_\mathrm{L})$. In normal-incidence reflection, the linear polarization will rotate by the so-called Polar Kerr angle: $\theta_\mathrm{K} = -\mathcal{I}m\{(\tilde{n}_\mathrm{L}-\tilde{n}_\mathrm{R})/(\tilde{n}_\mathrm{L}\tilde{n}_\mathrm{R} - 1)\}$ \cite{argyres}, which also suggests that the interference of two circularly polarized beams with opposite circular polarizations, reflected from a TRSB sample, contains a phase shift which is proportional to $\theta_\mathrm{K}$. Using the zero-area Sagnac Interferometer we can extract this phase shift with shot-noise limited sensitivity at optical power as low as 3 $\mu$W, while rejecting all reciprocal effects which do not break TRS \cite{xia2,kapitulnik}.  In addition, since a circularly polarized light impinging on a birefringent sample becomes elliptical, by monitoring the amount of ellipticity of the reflected beam through measurement of the second harmonic signal of the interferometer, $V_{2\omega}$ \cite{xia2,kapitulnik}, we could obtain the change in birefringence of the sample  \cite{hovo1}. Figure~\ref{structural} shows a scan of $V_{2\omega}$ as a function of temperature. As we cool the sample down, the signal first shows a very weak temperature dependence as expected from a tetragonal phase \cite{hovo1}.  At $T_{\mathrm{HT}}=230.8$~K the amplitude changes indicating the emergence of linear birefringence in the sample. Clearly this change is expected when the sample undergoes a phase transition to the LTO phase.  The birefringence continues to change as the temperature is lowered. Cooling the sample further, the birefringence of the sample exhibits an abrupt drop at $T_{\mathrm{LT}}=53.7$~K, coincidence with the LTO to LTT transition, which was previously shown to be first-order \cite{axe,billinge}. Indeed, the sharpness of the transition and its hysteretic behavior (see inset of Fig.~\ref{structural}) confirm that we correctly located this transition with our optical measurement. Furthermore, the weak temperature dependence below $T_{\mathrm{LT}}$ is again expected from this tetragonal phase \cite{hovo1}. 

Having established the fact that our apparatus indeed observes the relevant structural phase transitions, we turn back to Fig.~\ref{features}. Concentrating on the low temperature part of the data, we notice that the Kerr effect is very small (practically zero) above $T_{\mathrm{LT}}$. However, at $T_{\mathrm{LT}}$ the Kerr signal starts to increase rapidly (we show the birefringence signal for reference to mark the location of the transition). The Kerr signal further experiences a weak kink at $T_{\mathrm{CO}}\sim50$~K, which when compared to Tranquada {\it et al.} \cite{tranquada1} coincides with the CO transition. The signal continues to rise to a maximum that in different samples is found between 25~K and 40~K, which we will identify as the spin-order transition $T_{\mathrm{SO}}$ \cite{tranquada1}. We further find that the stronger the Kerr signal (in different samples), the closer to 40~K is the SO peak. The signal levels below $T_{\mathrm{SC}}\sim25$~K, which marks the temperature where in-plane superconductivity is established \cite{qli1,tranquada1}.  Susceptibility measured on the same sample (shown in Fig.~\ref{features}) is identical to previously measured susceptibility on \LBBCO~\cite{hucker1}, and confirms the location of the  transition temperatures identified with the Kerr effect. It is important to note a few interesting observations: i) The Kerr effect onsets at a structural phase transition, which in principle does not break TRS, ii) The Kerr signal starts to decrease below the spin-order transition, suggesting that if the Kerr effect originates from the local Cu-spins, they order in some antiferromagnetic fashion below $T_{\mathrm{SO}}$, iii) there is a finite Kerr effect in the superconducting state below $T_{\mathrm{SC}}$. 

\begin{figure}[h]
\begin{center}
\includegraphics[width=1.0 \columnwidth]{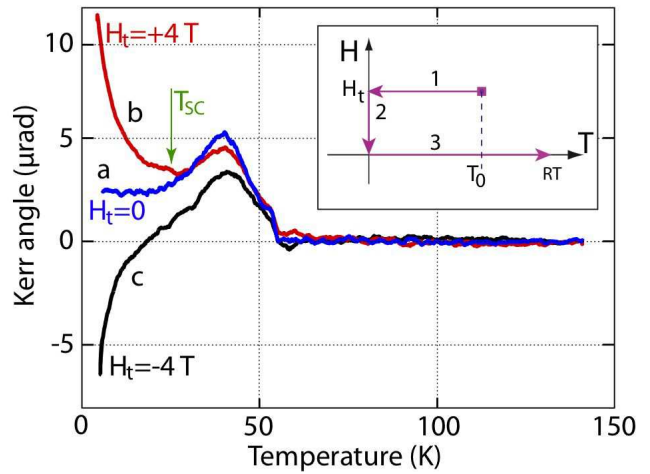}
\end{center}
\vspace{-5mm}
\caption{Training effect for a LBCO crystal. Inset shows the schedule of training: the field was turned on to $H_{\mathrm{t}}$ at $T_0$, the sample was then cooled in a field to 4~K, the field was turned off at 5~K, and the sample was measured at zero field while warmed up.  (a) Zero-field-cool: $T_0=$ Room Temperature (RT) and $H_{\mathrm{t}}=0$; (b) succeeded (a) where $T_0=140$~K and $H_{\mathrm{t}}=+4$~T;  (c) succeeded (b) where $T_0=140$~K and $H_{\mathrm{t}}=-4$~T. Vertical dashed line is the low field irreversibility temperature, indicating that when turning the field to zero at low temperatures, flux is trapped as is evident from the response below $\sim25$~K in (b) and (c). }
\label{training1}
\end{figure}

The first point, which is reminiscent of our previous studies of Kerr effect in \YBCO~\cite{xia1}, is the most puzzling one since there is no obvious reason why TRSB will coincide with the structural phase transition. Thus, to obtain further insight into the relation between the two effects, we attempted to train the sign of the TRSB with a  magnetic field  applied along the $c$-axis through $T_{\mathrm{LT}}$. If indeed TRSB is broken only below $T_{\mathrm{LT}}$,  the applied field would reverse the sign of the Kerr effect. Fig.~\ref{training1} is an example of such an attempt.  For calibration, the sample was first cooled in zero field to 5~K, and both, Kerr effect and birefringence were measured while warming up to 140~K. A magnetic field of +4~T  was then applied at 140~K, and the sample was cooled in that field to 5~K. The magnetic field was removed at 5~K and the sample was measured when warming up at earth field ($<0.4$~Oe).  At the end of this cycle a magnetic field of $-4$~T was applied and the sample was cooled in that field to 5~K, where again, the magnetic field was removed, and the sample was measured when warming up at earth field.

A close examination of the training results show that a field of 4~T, applied at 140~K could not fully reverse the sign of the Kerr effect, despite being applied much above the onset temperature of the Kerr signal. However, changing the cooling field from $+4$~T to $-4$~T resulted in a reduction of about 20$\%$ of the signal. A strong effect of the field is seen, though, when we look at the Kerr response below $T_{\mathrm{SC}}$. Here we believe that trapped vortices provide the signal which correlates with the direction and magnitude of the applied magnetic field (cooling at lower field shows smaller deviation of the Kerr signal from the flat zero field cooled one.) Subsequent studies in which we warmed the sample to room temperature (RT) to apply the $\pm4$~T did not change much the magnitude of the measured Kerr signal.

\begin{figure}[h]
\begin{center}
\includegraphics[width=1.0 \columnwidth]{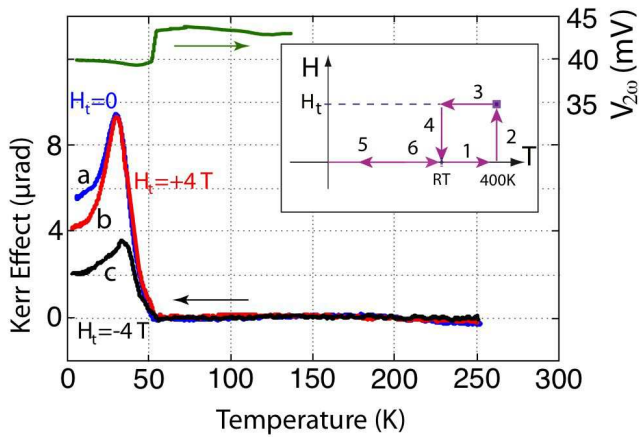}
\end{center}
\vspace{-5mm}
\caption{Training effect for a LBCO crystal at high temperatures. Sample was first cooled in zero field to 5~K, and was measured warming up to RT (a). Sample was then warmed up to 400~K , then cooled down to RT in a field of $+4$ T, followed by cooling in zero field to 5~K, and then measured warming up to RT (b). The same procedure was repeated but with a field of $-4$~T (c). We also show the second-harmonic that is proportional to the birefringence of the sample. } 
\label{training2}
\end{figure}

The above analysis implies that TRS has been broken much above RT. Thus, we set up the experiment described in Fig.~\ref{training2}. Here we warmed the sample to 400~K, applied a magnetic field of $\pm4$~T, cooled to RT, and transferred the sample to the Kerr apparatus to be measured down to 5~K at zero magnetic field (remnant $<3$~mOe.)  While the zero field and $+4$~T are basically identical, there is a very strong reduction, of about 70$\%$ of the signal, when we applied a field of $-4$~T, opposing to the sign of the Kerr effect. This strongly suggests that if we could apply a higher field, or apply the field at much higher temperatures, we could reverse the sign of the effect. Thus, similar to our conclusion in the study of \YBCO~\cite{xia1}, we speculate here as well that TRS was broken at much higher temperatures. Note that since we did not cool the sample in a field below $T_{\mathrm{SC}}$, no vortex effect is observed in this experiment. 

Evidence for TRSB  in \LBBCO~was recently reported by Li {\it et al.} \cite{luli}, showing the existence of a large, anomalous Nernst signal below $T_{\mathrm{LT}}\sim T_{\mathrm{CO}}$ that is symmetric in magnetic field $H$, and remains finite as $H \rightarrow 0$. However, despite the sharp onset of the effect, and the tracking of the various characteristic temperatures with the Nernst signal, attempts to force the system into a particular ``sign" of TRSB failed. In particular, Li {\it et al.} were searching for an open-loop hysteretic dependence  between $+14$ and $-14$ T both above and below $T_{\mathrm{SC}}$ but could not alter the sign of the effect. Similar to our Kerr effect results, these observations also suggest that TRS has been broken at a much higher temperature.

Earlier studies of Hall effect \cite{hwang}, susceptibility \cite{yoshizaki} and Knight shift \cite{zheng} in the sister material \LSCO,  indicate a characteristic temperature that for $x\approx 1/8$ is around $T^* \approx 530$~K, where \LBCO is expected to show similar behavior. Recent studies of the magnetic excitations in \LBBCO~provide strong evidence for dynamic stripes that persist to high energies \cite{gxu}.  It is therefore reasonable to speculate that TRS breaking occurs on this scale as well, possibly at a temperature where dynamic stripes develop substantial correlations. Below $T_{\mathrm{LT}}\sim T_{\mathrm{CO}}\sim 54$~K, where CO correlations appear \cite{tranquada1,wilkins}, accompanied by static magnetic order  \cite{kumagai,luke,sera,arai,hucker1}, and strong elastic spin scattering \cite{tranquada1}. At this point, while the Kerr signal detects a $k=0$ component of any relevant order parameter, and thus cannot by itself give any clues concerning the character of the putative density wave order that was used to explain a variety of novel results in the cuprates, including the Fermi pockets detected in quantum oscillation experiments \cite{millis_norman}, the present results do, however, support the previous conjecture that the Kerr signal rides on top of a charge ordering transition, thus, indicating strong changes in spin-orbit interaction through that transition, and/or change in crystal symmetry, e.g. allowing for a possible magnetoelectric effect \cite{orenstein}. CO is then followed by  spin-ordering within the stripe phase at $T_{\mathrm{SO}}\sim 40$~K, below which ($\lesssim 35$~K) two-dimensional superconducting fluctuations are found \cite{qli1}.  Global superconductivity appears only at much lower temperature, $T_{\mathrm{c}}\sim 4$~K, which led Berg {\it et al.} \cite{berg1,berg2,berg3} to propose a unique pair density wave (PDW) scenario to explain the dynamical decoupling of the layers above $T_{\mathrm{c}}$. An important consequence of this theory is that, in the presence of weak quenched disorder (see e.g. \cite{tranquada1}), the superconducting phase gives way to a glassy striped phase that is characterized by its spontaneous TRSB. The fact that we observe a finite Kerr effect at low temperatures, which seems to be smoothly obtained from high temperatures and is position dependent, could be evidence for the PDW phase \cite{berg3}.  Within this model,  the zero-field Nernst signal found  by Li {\it et al.} \cite{luli}, was suggested to be due to an array of 2D vortices spontaneously nucleated below $\sim T_{\mathrm{CO}}$ \cite{berg1}.  Thus, as we have shown that vortices contribute to the Kerr effect (see Fig.~\ref{training1}), we can estimate that the maximum Kerr signal is equivalent to the signal of vortices trapped after cooling in $\sim4$ T field, which can give us an estimate of the internal field of the TRSB. 

In conclusion, our observations clearly highlight the connection between magnetic and TRSB effects and charge- and spin-ordering in \LBCO\;and possibly in all HTSC. We see that charge order appears to be the leading order that both competes and coexists with the bulk superconductivity. 

\acknowledgments
Discussions with Alexander Fried and Steve Kivelson are greatly appreciated.  This work was supported by the Department of Energy Grant DE-AC02-76SF00515.

\end{document}